\documentclass[a4paper,11pt]{article}
\usepackage{paper} 
\newcommand{\be}{\begin{equation}}
\newcommand{\ee}{\end{equation}}

\usepackage{color}

\title{\boldmath  Black hole surrounded by the pseudo-isothermal dark matter halo}

\author[1]{Yi Yang,}
\emailAdd{yangyigz@yeah.net}
\affiliation[1]{School of Mathematics and Statistics, \\
Guizhou University of Finance and Economics, Guiyang, 550025, China}

\author[2]{Dong Liu,}
\emailAdd{dongliuvv@yeah.net}
\affiliation[2]{College of Physics, Guizhou University, Guiyang, 550025, China}

\author[3]{Ali \"Ovg\"un,}
\emailAdd{ali.ovgun@emu.edu.tr}
\affiliation[3]{Physics Department, Eastern Mediterranean University, Famagusta, 99628 North Cyprus via Mersin 10, Turkey}

\author[4,5]{Gaetano Lambiase,}
\emailAdd{lambiase@sa.infn.it}
\affiliation[4]{Dipartimento di Fisica ``E.R Caianiello'', Università degli Studi di Salerno, Via Giovanni Paolo II, 132 - 84084 Fisciano (SA), Italy}
\affiliation[5]{Istituto Nazionale di Fisica Nucleare - Gruppo Collegato di Salerno - Sezione di Napoli, Via Giovanni Paolo II, 132 - 84084 Fisciano (SA), Italy}

\author[2]{Zheng-Wen Long}
\emailAdd{zwlong@gzu.edu.cn}
\affiliation[2]{College of Physics, Guizhou University, Guiyang, 550025, China}

\abstract{In this paper, we obtain a new spherically symmetric black hole surrounded by the pseudo-isothermal dark matter halo. Furthermore, to explore the effects of the pseudo-isothermal halo profile on a rotating black hole at the M87 galactic center, we derive a rotating black hole solution encompassed by the pseudo-isothermal halo by using the Newman-Janis method. Our investigation focuses on the impact of the pseudo-isothermal halo on the black hole event horizon, time-like and null orbits, as well as the black hole shadow. We find that as the spin parameter $a$ increases, the interval between the inner event horizon and the outer event horizon of the rotating black hole surrounded by the pseudo-isothermal halo in M87 diminishes. This leads to the formation of an extreme black hole. The presence of dark matter, however, has minimal effect on the event horizon. Moreover, in the M87 as the spin parameter $a$ increases, the black hole shadow deviates increasingly from a standard circle, with larger spin parameters causing more pronounced distortion relative to the standard circle. Surprisingly, we observe that the dark matter density has very little influence on the shadow of the black hole surrounded by the pseudo-isothermal halo in the M87. This study contributes to a deeper understanding of black hole structures and the role of dark matter in the universe.}

\arxivnumber{}

\begin{document}
\maketitle
\flushbottom

\section{Introduction}\label{sec:intro}
In 1915, Einstein put forth the groundbreaking theory of general relativity (GR) in the realm of physics, where he linked the gravitational field to the curvature of spacetime. Within this theory emerged the concept of black holes (BH), extraordinary celestial objects in the universe. Their existence was not observed directly but rather predicted based on the principles of general relativity. In 1916, Schwarzschild \cite{Schwarzschild:1916uq} became the first to solve the field equations of general relativity, resulting in a static spherically symmetric vacuum solution that described a non-rotating black hole with mass alone. Later, in 1963, Kerr derived a rotating, steady-state axisymmetric vacuum black hole solution \cite{Kerr:1963ud}. However, within classical gravitational theory, a singularity exists inside the black hole \cite{Joshi:2000fk,Tsukamoto:2020bjm,Barrientos:2022avi,Ovgun:2020yuv}.

On the other hand, the true nature of dark matter is one of the most profound mysteries in astrophysics and cosmology. The widely accepted $\Lambda$ CDM model has been remarkably successful in explaining the dynamics of the large-scale Universe, suggesting that about 26.8$\%$ of our Universe is composed of dark matter, making up a significant 85$\%$  of its total mass \cite{Planck:2013oqw}. In the early universe, the dark matter halo has been suggested as a potential factor in understanding this phenomenon \cite{Balberg:2001qg,Balberg:2002ue}. Although numerical simulations have been used to analyze the growth of black holes within dark matter halos over time, obtaining an analytical form for black holes surrounded by dark matter halo remains challenging. This difficulty arises from the unknown and uncertain behavior of dark matter particles when they interact with black holes.

The presence of a supermassive black hole at the center of a galaxy can significantly enhance the dark matter density, leading to a phenomenon called the ``Spike" \cite{Gondolo:1999ef,Sadeghian:2013laa,Fields:2014pia}. However, the  Navarro-Frenk-White (NFW) density profile exhibits a ``cusp" problem \cite{de_Blok_2010}, which contradicts observations showing a rather flat density profile. Different dark matter models, such as scalar field dark matter, modified Newtonian dynamics dark matter, and warm dark matter, do not produce the ``cusp" in small scales. It remains uncertain whether the ``Spike" and ``Cusp" manifest in the galactic center.

These challenges serve as motivation for our study, where we examine black holes (spherically symmetric and rotating) in dark matter halos under stationary conditions. By exploring the results obtained, we can investigate various dynamic processes occurring near the black hole, and the energy density of dark matter under relativistic conditions. This research seeks to shed light on the behavior of dark matter in the vicinity of black holes and provides insights into potential solutions for the observed discrepancies between theoretical predictions and observational data. One notable method for extending Schwarzschild-like black holes to Kerr-like black holes is the Newman-Janis (NJ) algorithm \cite{Newman:1965tw,Azreg-Ainou:2014nra,Azreg-Ainou:2014pra}, which has proven to be highly suitable for this purpose. Additionally, reference \cite{Shaikh:2021yux} also contributes to the topic.

In 2019, the Event Horizon Telescope (EHT) released the first image of a black hole, capturing the shadow of the supermassive black hole M87 at the center of the Virgo elliptical galaxy \cite{EventHorizonTelescope:2019dse}. Moreover, in 2022, EHT released an image of Sagittarius A*, the supermassive black hole at the center of the Milky Way \cite{EventHorizonTelescope:2022apq}. The information revealed in these black hole photos has provided valuable insights into the shadow, jet, and accretion processes of black holes. As a result of these groundbreaking discoveries, the phenomenon of black hole shadows has attracted significant attention in the scientific literature. Numerous studies have been conducted to delve into this intriguing aspect \cite{Capozziello:2023tbo,Feng:2022evy,Banerjee2019nnj,Banerjee:2022jog,Falcke:1999pj,Xu:2018mkl,Hou:2018bar,Hou:2018avu,Okyay:2021nnh,Yang:2023agi,Kumaran:2023brp,Lambiase:2023hng,Uniyal:2022vdu,Uniyal:2023inx,Ovgun:2023ego,Atamurotov:2022knb,Kumaran:2022soh,Pantig:2022qak,Mustafa:2022xod,Rayimbaev:2022hca,Pantig:2022ely,Chakhchi:2022fls,Kuang:2022xjp,Cimdiker:2021cpz,Symmergent-bh4,
Tsukamoto:2014tja,Tsukamoto:2017fxq,Shaikh:2019fpu,Shaikh:2018kfv,Kasuya:2021cpk,Ovgun:2018tua,Fathi:2022ntj,Belhaj:2020kwv,Belhaj:2020okh,Vagnozzi:2022moj,Chen:2022nbb,Roy:2021uye,Vagnozzi:2020quf,Bambi:2008jg,Bambi:2019tjh,Bambi:2010hf,Zakharov:2014lqa,Zakharov:2005ek,Zakharov:2011zz,Zakharov:2018syb,Zakharov:2021gbg,KumarWalia:2022aop,Konoplya:2021slg,Khodadi:2021gbc,Khodadi:2022pqh,Pantig:2022sjb,Pantig:2022toh,Pantig:2021zqe,Pantig:2022whj,Nampalliwar:2021tyz,Xavier:2023exm,Parbin:2023zik,Mustafa:2023vvt,Capozziello:2023rfv,Qin:2023nog,Mustafa:2022fxn,Pugliese:2022oes,Konoplya:2022hbl,Konoplya:2019sns,Bhandari:2021dsh,Zhang:2021bdr,Saurabh:2020zqg,Contreras:2020kgy,Contreras:2019cmf,Contreras:2019nih,Herdeiro:2021lwl,Sengo:2022jif,Junior:2021svb,Junior:2021dyw,Lima:2021las,Cunha:2018cof,Cunha:2018gql,Cunha:2015yba,Guerrero:2021ues,Guerrero:2021pxt,Sharif:2016znp,Amir:2018pcu,Khodadi:2022ulo,Meng:2023wgi,Wang:2023vcv,Li:2023djs,Kuang:2022ojj,Meng:2022kjs,Anjum:2023axh,Islam:2022wck,Ghosh:2022kit,Ghosh:2020spb,Abdujabbarov:2016hnw,Kumar:2020owy,Papnoi:2014aaa,Kumar:2018ple,Toshmatov:2014nya,Atamurotov:2013sca,Abdujabbarov:2015xqa,Atamurotov:2015nra}.
In Ref. \cite{Tang2022uwi} Tang and Xu presented a spacetime metric for a rotating short-hairy black hole, and studied the influence of short-hairy on the shadow of black holes. These studies have contributed significantly to our understanding of black hole properties and their impact on the surrounding environment. In our study, we will utilize the black hole shadow to gain valuable insights into the effects of pseudo-isothermal halo (PIH) profile \cite{Begeman:1991iy} on a rotating black hole geometry. To achieve this, we will primarily rely on the formalism developed by Xu et al.  \cite{Xu:2018wow,Xu:2021dkv} to obtain the combined dark matter and black hole geometries. By employing these techniques, we aim to elucidate the impact of pseudo-isothermal halo profile on a rotating black hole, specifically in the context of their observable shadows. This investigation can contribute to a deeper understanding of the interaction between black holes and dark matter and may offer valuable clues for unraveling the enigmatic nature of dark matter through astrophysical observations.

In our research, we aim to explore the shadows of rotating black holes surrounded by dark matter and analyze their unique characteristics. By doing so, we hope to provide valuable insights and directions for the experimental detection of this particular spacetime configuration. This investigation will contribute to our understanding of black hole properties and enhance our ability to interpret the observations of black hole shadows in astrophysical contexts.
This work is organized as follows. In Sec. \ref{PIandF}, we give a pure dark matter spacetime metric by considering the pseudo-isothermal halo profile, and solve the Einstein field equations to obtain the spherically symmetric BH metric surrounded by the pseudo-isothermal halo. In Sec. \ref{rbh_pi} we derive the rotating black hole surrounded by the pseudo-isothermal halo profile derived by using the Newman-Janis method. In Sec. \ref{HJ-eq}, we obtain the geodesic equation by solving the Hamilton-Jacobi equation in the rotating black hole. In Sec. \ref{shape_shadow}, we present the shadow images of a rotating black hole and analyze the influence of spin and dark matter parameters on the shadow. Sec. \ref{sec:summary} provides our main conclusion of this work.

\section{Spherically symmetric BH metric surrounded by the PIH}\label{PIandF}
In this work, we study the pseudo-isothermal halo profile \cite{Begeman:1991iy}. The dark matter  density profile reads
\begin{equation}
\rho_{PI}(r)=\rho_0\left[1+\left(\frac{r}{r_{\mathrm{0}}}\right)^2\right]^{-1}.
\end{equation}
where $\rho_0$ denotes the central halo density and $r_0$ denotes the halo core radius. In order to obtain a spherically symmetric black hole surrounded by dark matter, we first need to calculate the mass distribution of the dark matter halo. According to the above dark matter halo profile, we can obtain the mass profile \cite{Matos:2003nb})
\begin{equation}
M_{\text {PI }}=4 \pi \int_0^r \rho_{\text {PI}}\left(r^{\prime}\right) r^{\prime 2} d r^{\prime}=4 \pi  \rho _0 r_0^2 \left[r-r_0\arctan
  \left(\frac{r}{r_0}\right)\right].
\end{equation}
In the spherically symmetric spacetime, we can find the tangential velocity of the test particle moving in the dark matter halo determined by the mass distribution of the dark matter halo. Therefore, we obtain the tangential velocity (in units $G=1,c=1$)
\begin{equation}\label{vpi}
V_{\text {PI}}=\sqrt{\frac{M_{\text {PI }}}{r}}=\sqrt{\frac{ 4\pi  \rho _0 r_0^2 \left[r-r_0\arctan
  \left(\frac{r}{r_0}\right)\right]}{r}}.
\end{equation}
The spherically symmetric spacetime line elements describing pure dark matter halo can be written as (using the method introduced in Ref. \cite{Matos:2003nb} and later used by Xu et al. in Ref.\cite{Xu:2018wow})
\begin{equation}\label{pure_g}
d s^2=-\mathcal{A}(r) d t^2+\mathcal{B}(r)^{-1} d r^2+r^2\left(d \theta^2+\sin ^2 \theta d \phi^2\right),
\end{equation}
where the $\mathcal{A}(r)$ represents the redshift functions, and $\mathcal{B}(r)$  represents the shape functions. The tangential velocity is closely related to the redshift function $\mathcal{A}(r)$ of the spacetime metric, which reads
\begin{equation}
V_{PI}^2=r \frac{d \ln \sqrt{\mathcal{A}(r)}}{d r} .
\end{equation}
It should be noted that in our work, we consider the case of $\mathcal{A}(r)=\mathcal{B}(r)$. Substituting equation (\ref{vpi}) into the above equation, we can obtain the analytical expression of the redshift function and the shape function in the pure dark matter metric
\begin{equation}\label{fPI}
\mathcal{A}_{PI}(r)=\mathcal{B}_{PI}(r)=\left(r_0^2+r^2\right){}^{4 \pi  \rho _0 r_0^2} \text{exp}\left[\frac{8 \pi  \rho _0 r_0^3  \arctan
   \left(\frac{r}{r_0}\right)}{r}\right].
\end{equation}

Now, we will solve the Einstein field equations to obtain the spherically symmetric BH metric surrounded by the pseudo-isothermal halo. As proposed by Xu et al. \cite{Xu:2018wow}, the Einstein field equation for the pure pseudo-isothermal dark matter spacetime needs to be solved, which can be written as
\begin{equation}
R_{\mu \nu}-\frac{1}{2} g_{\mu \nu} R=\kappa^2 T_{\mu \nu}(\mathrm{PIH}),
\end{equation}
where $g_{\mu \nu}$ denotes the metric tensor of the pure pseudo-isothermal dark matter spacetime, $R_{\mu \nu}$ and $R$ denotes the Ricci tensor and  Ricci
scalar, respectively.  The
energy-momentum tensor of the pure pseudo-isothermal dark matter spacetime can be written as $T^\nu{}_\mu=g^{\nu \sigma} T_{\mu \sigma}=\operatorname{diag}\left[-\rho, p_r, p, p\right]$. Therefore, we can obtain
\begin{equation}
\begin{aligned}
& \kappa^2 T^t{}_t(\mathrm{PIH})=\mathcal{B}_{PI}(r)\left(\frac{1}{r} \frac{\mathcal{B}_{PI}^{\prime}(r)}{\mathcal{B}_{PI}(r)}+\frac{1}{r^2}\right)-\frac{1}{r^2}, \\
& \kappa^2 T^r{}_r(\mathrm{PIH})=\mathcal{B}_{PI}(r)\left(\frac{1}{r^2}+\frac{1}{r} \frac{\mathcal{A}_{PI}^{\prime}(r)}{\mathcal{A}_{PI}(r)}\right)-\frac{1}{r^2}, \\
& \kappa^2 T^\theta{}_\theta(\mathrm{PIH})=\kappa^2 T^\phi{}_\phi(\mathrm{PIH})=\frac{1}{2} \mathcal{B}_{PI}(r)\times\\
&\left[\frac{\mathcal{A}_{PI}^{\prime \prime}(r) \mathcal{A}_{PI}(r)-\mathcal{A}_{PI}^{\prime 2}(r)}{\mathcal{A}_{PI}^2(r)}+\frac{\mathcal{A}_{PI}^{\prime 2}(r)}{2 \mathcal{A}_{PI}^2(r)}+\frac{1}{r}\left(\frac{\mathcal{A}_{PI}^{\prime}(r)}{\mathcal{A}_{PI}(r)}+\frac{\mathcal{B}_{PI}^{\prime}(r)}{\mathcal{B}_{PI}(r)}\right)+\frac{\mathcal{A}_{PI}^{\prime}(r) \mathcal{B}_{PI}^{\prime}(r)}{2 \mathcal{A}_{PI}(r) \mathcal{B}_{PI}(r)}\right].
\end{aligned}
\end{equation}
If consider a black hole surrounded by the PIH, the energy-momentum tensor becomes $T^\nu{}_\mu=T^\nu{}_\mu(\mathrm{BH})+T^\nu{}_\mu(\mathrm{PIH})$.
According to GR, a Schwarzschild black hole is a vacuum solution, which satisfies the condition $T^\nu{}_\mu(\mathrm{BH})=0$. If we consider a Schwarzschild-like BH surrounded by the PIH, we can only consider the energy-momentum tensor of the pure PIH.
Therefore, we assume the  black hole metric surrounded by the PIH has the following form
\be\label{newm}
d s^2=-[\mathcal{A}_{PI}(r)+X_1(r)] d t^2+\frac{1}{\mathcal{B}_{PI}(r)+X_2(r)} d r^2+r^2\left(d \theta^2+\sin ^2 \theta d \phi^2\right),
\ee
where $\mathcal{A}_{PI}(r)$ and $\mathcal{B}_{PI}(r)$ denote the metric coefficients of the pure PIH and the unknown functions $X_1(r)$ and $X_2(r)$ are determined by the black hole parameters and PIH parameters. As a consequence, the Einstein field equations can be cast in the form
\be
R_{\mu \nu}-\frac{1}{2} g_{\mu \nu} R=\kappa^2\left[T_{\mu \nu}(\mathrm{BH})+T_{\mu \nu}(\mathrm{PIH})\right].
\ee
Substituting the new black hole metric (\ref{newm}) into the above Einstein field equation, we can obtain
\begin{subequations}
\begin{align}
& \left[\mathcal{B}_{PI}(r)+X_2(r)\right]\left[\frac{1}{r^2}+\frac{1}{r} \frac{\mathcal{B}_{PI}^{\prime}(r)+X_2 ^{\prime}(r)}{\mathcal{B}_{PI}(r)+X_2(r)}\right]=\mathcal{B}_{PI}(r)\left[\frac{1}{r^2}+\frac{1}{r} \frac{\mathcal{B}_{PI}^{\prime}(r)}{\mathcal{B}_{PI}(r)}\right], \\
& \left[\mathcal{B}_{PI}(r)+X_2(r)\right]\left[\frac{1}{r^2}+\frac{1}{r} \frac{\mathcal{A}_{PI}^{\prime}(r)+X_1 ^{\prime}(r)}{\mathcal{A}_{PI}(r)+X_1(r)}\right]=\mathcal{B}_{PI}(r)\left[\frac{1}{r^2}+\frac{1}{r} \frac{\mathcal{A}_{PI}^{\prime}(r)}{\mathcal{A}_{PI}(r)}\right].
\end{align}
\end{subequations}
Using the Schwarzschild black hole as the boundary condition, we can obtain the analytical solution of the above two differential equations
\begin{equation}
\begin{aligned}
& X_2(r)=-\frac{2 M}{r},\\
&X_1(r)=\exp \left\{\int \frac{\mathcal{B}_{PI}(r)}{\mathcal{B}_{PI}(r)+X_2(r)}\left[\frac{1}{r}+\frac{\mathcal{A}_{PI}^{\prime}(r)}{\mathcal{A}_{PI}(r)}\right] d r-\frac{1}{r} d r\right\}-\mathcal{A}_{PI}(r).
\end{aligned}
\end{equation}
For the pseudo-isothermal halo, if $\mathcal{A}_{PI}(r)=\mathcal{B}_{PI}(r)$, one can find that $X_1(r)=X_2(r)=-2 M/ r$. The black hole surrounded by the pseudo-isothermal halo can be hence written as
\begin{equation}
d s^2=-Q_1(r) d t^2+\frac{1}{Q_2(r)}d r^2+\mathcal{R}(r)\left(d \theta^2+\sin ^2 \theta d \phi^2\right),
\end{equation}
where $\mathcal{R}(r)=r^2$, and $Q_1=X_1(r)+\mathcal{A}_{PI}(r)=Q_2=X_2(r)+\mathcal{B}_{PI}(r)$, whose explicit expressions are
\be
Q_1(r)=Q_2(r)=\left(r_0^2+r^2\right){}^{4 \pi  \rho _0 r_0^2} \text{exp}\left[\frac{8 \pi  \rho _0 r_0^3  \arctan
   \left(\frac{r}{r_0}\right)}{r}\right]-\frac{2 M}{r}.
\ee

\section{Rotating black hole surrounded by the pseudo-isothermal halo}\label{rbh_pi}
The black hole photos taken by EHT and the detection of gravitational wave events indicate that the black holes in the actual universe are Kerr-like black holes, i.e., they usually have spins. Therefore, we will introduce spin into the spherically symmetric black hole spacetime. To achieve this, we will use the Newman-Janis algorithm, which allows us to introduce the spin into the black hole surrounded by the pseudo-isothermal halo. In the NJ method, in order to obtain the spacetime metric of a rotating black hole surrounded by the pseudo-isothermal halo, it is necessary to transform the spherically symmetric metric from Boyer-Lindquist to Eddington-Finkelstein coordinates. The relationship between the Boyer-Lindquist coordinates $(t,r,\theta,\phi)$ and the advanced null coordinate $(u,r,\theta,\phi)$ is
\begin{equation}
\mathrm{d} u=dt-dr_*=\mathrm{d} t-\frac{\mathrm{d} r}{Q_1(r)} .
\end{equation}
The contravariant metric tensor $g^{\mu \nu}$ for a spherically symmetric metric can be given by the four basis vectors ($l^\mu, n^\mu, m^\mu$, $\bar{m}^\mu$)
\begin{equation}\label{guu}
g^{\mu \nu}=-l^\mu n^\nu-l^\nu n^\mu+m^\mu \bar{m}^\nu+m^\nu \bar{m}^\mu,
\end{equation}
where the four basis vectors satisfy the orthogonality conditions and the modulus of the basis vector is one, i.e.
\begin{equation}
\begin{aligned}
l^\mu l_\mu & =n^\mu n_\mu=m^\mu m_\mu=\bar{m}^\mu \bar{m}_\mu=0, \\
l^\mu m_\mu & =l^\mu \bar{m}_\mu=n^\mu m_\mu=n^\mu \bar{m}_\mu=0, \\
-l^\mu n_\mu & =m^\mu \bar{m}_\mu=1 .
\end{aligned}
\end{equation}
For the metric of a rotating black hole surrounded by the pseudo-isothermal halo, the basis vectors are
\begin{equation}
\begin{aligned}
l^\mu & =\delta_r^\mu, \\
n^\mu & =\sqrt{\frac{Q_2(r)}{Q_1(r)}} \delta_\mu^\mu-\frac{Q_1(r)}{2} \delta_r^\mu, \\
m^\mu & =\frac{1}{\sqrt{2} \mathcal{R}(r)} \delta_\theta^\mu+\frac{i}{\sqrt{2\mathcal{R}(r)}  \sin \theta} \delta_\phi^\mu, \\
\bar{m}^\mu & =\frac{1}{\sqrt{2} \mathcal{R}(r)} \delta_\theta^\mu-\frac{i}{\sqrt{2\mathcal{R}(r)}  \sin \theta} \delta_\phi^\mu.
\end{aligned}
\end{equation}
According to the NJ method, the spacetime coordinates for different observers satisfy the following complex transformation
\begin{equation}
u \rightarrow u-i a \cos \theta, \quad r \rightarrow r-i a \cos \theta.
\end{equation}
Using this transformation, the metric coefficient will include spin, i.e., the metric coefficient becomes a function of $(r,\theta,a)$. According to the above complex transformation, the metric functions become:
$Q_1(r) \rightarrow \mathcal{W}(r, \theta, a), Q_2(r) \rightarrow \mathcal{Z}(r, \theta, a)$ and $\mathcal{R}(r) \rightarrow \Sigma(r, \theta, a)$. After considering this change, the null tetrad become

\begin{equation}
\begin{aligned}
l^\mu & =\delta_r^\mu, \\
n^\mu & =\sqrt{\frac{\mathcal{Z}}{\mathcal{W}}} \delta_\mu^\mu-\frac{\mathcal{Z}}{2} \delta_r^\mu, \\
m^\mu&=\frac{1}{\sqrt{2 \Sigma}}\left[\delta_\theta^\mu+i a \sin \theta\left(\delta_u^\mu-\delta_r^\mu\right)+\frac{i}{\sin \theta} \delta_\phi^\mu\right], \\
\bar{m}^\mu &=\frac{1}{\sqrt{2\Sigma}}\left[\delta_\theta^\mu-i a \sin \theta\left(\delta_u^\mu-\delta_r^\mu\right)-\frac{i}{\sin \theta} \delta_\phi^\mu\right] .
\end{aligned}
\end{equation}
Therefore, according to these null tetrads, the contravariant metric tensor of the rotating black hole surrounded by the pseudo-isothermal halo can be written as
\begin{equation}
g^{\mu \nu}=\left(\begin{array}{cccc}
\frac{a^2 \sin ^2 \theta}{\Sigma} & -\sqrt{\frac{\mathcal{Z}}{\mathcal{W}}}-\frac{a^2 \sin ^2 \theta}{\Sigma} & 0 & \frac{a}{\Sigma} \\
-\sqrt{\frac{\mathcal{Z}}{\mathcal{W}}}-\frac{a^2 \sin ^2 \theta}{\Sigma} & \mathcal{Z}+\frac{a^2 \sin ^2 \theta}{\Sigma} & 0 & -\frac{a}{\Sigma} \\
0 & 0 & \frac{1}{\Sigma} & 0 \\
\frac{a}{\Sigma} & -\frac{a}{\Sigma} & 0 & \frac{1}{\Sigma \sin ^2 \theta}
\end{array}\right) .
\end{equation}
Moreover, the covariant non-zero metric tensor components are given by
\begin{equation}
\begin{aligned}
& g_{u u}=-\mathcal{W}, \quad g_{u r}=-\sqrt{\frac{\mathcal{W}}{\mathcal{Z}}}, \\
&g_{u \phi}=a\sin ^2 \theta\left(\mathcal{W}-\sqrt{\frac{\mathcal{Z}}{\mathcal{W}}}\right),\\
& g_{r \phi}=a \sin ^2 \theta \sqrt{\frac{\mathcal{W}}{\mathcal{Z}}} , \quad g_{\theta \theta}=\Sigma, \\
& g_{\phi \phi}=\Sigma\sin ^2 \theta+a^2\left(2 \sqrt{\frac{\mathcal{Z}}{\mathcal{W}}}-\mathcal{\mathcal{W}}\right) \sin ^4 \theta.
\end{aligned}
\end{equation}
Then, we can obtain the rotating black hole surrounded by the pseudo-isothermal halo in Eddington-Finkelstein coordinates
\begin{equation}
\begin{aligned}
d s^2= & -\mathcal{W} d u^2-2 \sqrt{\frac{\mathcal{W}}{\mathcal{Z}}} d u d r+2 a \sin ^2 \theta\left(\mathcal{W}-\sqrt{\frac{\mathcal{Z}}{\mathcal{W}}}\right) d u d \phi+2 a \sin ^2 \theta \sqrt{\frac{\mathcal{W}}{\mathcal{Z}}} d r d \phi+\Sigma d \theta^2 \\
& +\left[\Sigma\sin ^2 \theta+a^2\left(2 \sqrt{\frac{\mathcal{Z}}{\mathcal{W}}}-\mathcal{\mathcal{W}}\right) \sin ^4 \theta\right] d \phi^2.
\end{aligned}
\end{equation}
One significant challenge in describing black holes is the presence of coordinate singularities, particularly at the event horizon. The Boyer-Lindquist coordinates are chosen to avoid singularities at the event horizon, making them regular there, which is crucial for understanding the physical properties of the black hole, especially when studying the behavior of objects and light near the horizon. Moreover, The Boyer-Lindquist coordinates are manifestly covariant under coordinate transformations, which means that the mathematical expressions for physical quantities maintain the same form regardless of the chosen coordinates.
The number of non-zero off-diagonal components of the metric is minimal in Boyer–Lindquist coordinates \cite{Erbin:2016lzq}.
Therefore, transforming the Eddington-Finkelstein coordinates back to Boyer-Lindquist coordinates is necessary. To this purpose, we use the following coordinate transformation
\begin{equation}
\mathrm{d} u=\mathrm{d} t+\mathcal{T}_1(r) \mathrm{d} r, \quad \mathrm{~d} \phi=\mathrm{d} \phi^\prime+\mathcal{T}_2(r) \mathrm{d} r,
\end{equation}
where the $\mathcal{T}_1(r)$ and $\mathcal{T}_2(r)$ take the form \cite{AA}
\begin{equation}
\mathcal{T}_1(r)=-\frac{\mathcal{U}(r)+a^2}{Q_1(r) Q_2(r)+a^2}, \quad \mathcal{T}_2(r)=-\frac{a}{Q_1(r) Q_2(r)+a^2},
\end{equation}
with
\begin{equation}
\mathcal{U}(r)=\sqrt{\frac{Q_2(r)}{Q_1(r)}} \mathcal{R}(r),
\end{equation}
and
\begin{equation}
\mathcal{W}(r, \theta)=\frac{\left[Q_2(r) \mathcal{R}(r)+a^2 \cos ^2 \theta\right] \Sigma}{\left[\mathcal{U}(r)+a^2 \cos ^2 \theta\right]^2}, \quad \mathcal{Z}(r, \theta)=\frac{Q_2(r) \mathcal{R}(r)+a^2 \cos ^2 \theta}{\Sigma} .
\end{equation}
Therefore, the spacetime metric of the rotating black hole surrounded by the pseudo-isothermal halo can be given by
\be
\begin{aligned}
d s^2= & -\left(1-\frac{r^2-\mathcal{A}_{PI}(r) r^2+2 M r}{\Sigma}\right) d t^2+\frac{\Sigma}{\Delta} d r^2-2 a \sin ^2 \theta\left(\frac{r^2-\mathcal{A}_{PI}(r)r^2+2 M r}{\Sigma}\right) d t d \phi \\
& +\Sigma d \theta^2+\left(\left(a^2+r^2\right) \sin ^2 \theta+\frac{a^2 \sin ^4 \theta\left(r^2-\mathcal{A}_{PI}(r) r^2+2 M r\right)}{\Sigma}\right) d \phi^2,
\end{aligned}
\ee
where $\Sigma$ and $\Delta$ are defined by
\begin{subequations}
\begin{align}
\Sigma &=r^2+a^2 \cos ^2 \theta, \\
\Delta &=\left(r_0^2+r^2\right){}^{4 \pi  \rho _0 r_0^2} \text{exp}\left[\frac{8 \pi  \rho _0 r_0^3  \arctan
   \left(\frac{r}{r_0}\right)}{r}\right]-2 M r+a^2.
\label{DeltaGL}
\end{align}
\end{subequations}
For $a=0$, the rotating black hole surrounded by the pseudo-isothermal halo degenerates into the non-rotating black hole surrounded by the pseudo-isothermal halo, while
for $\rho_0=0$, the rotating black hole surrounded by the pseudo-isothermal halo degenerates into the Kerr black hole. The Schwarzschild black hole metric is recovered for $a=0,\rho_0=0$.

\begin{figure}[b!]
\begin{center}
\vspace{0.5cm}
\includegraphics[scale=0.6]{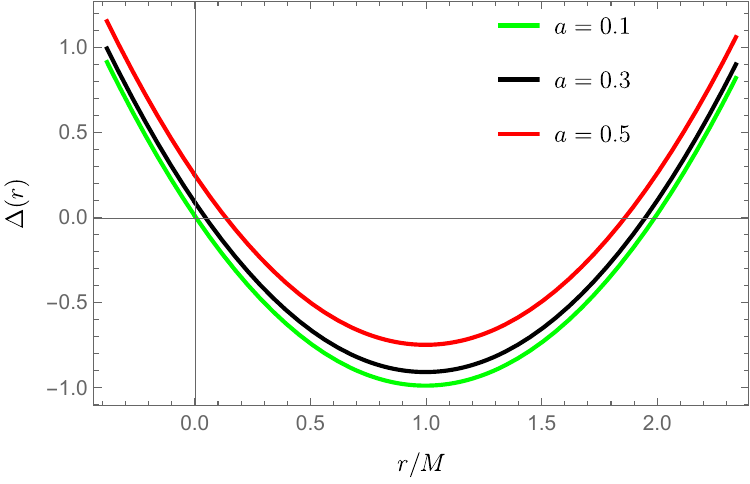}
\includegraphics[scale=0.6]{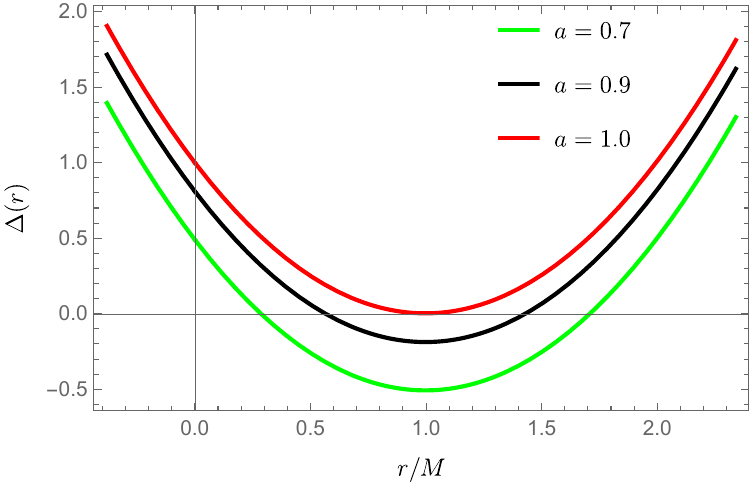}
\end{center}
\setlength{\abovecaptionskip}{-0.3cm}
\setlength{\belowcaptionskip}{0.9cm}
\caption{$\Delta(r)$ as the function of $r$  (see Eq. (\ref{DeltaGL})), with $\rho_0=6.9 \times 10^6 M_{\odot} / k p c^3, r_0=91.2 \mathrm{kpc}$. The three curves in the left panel correspond to the results of $a=0.1$, $a=0.3$ and $a=0.5$, respectively. The three curves in the right panel correspond to the results of $a=0.7$, $a=0.9$ and $a=1$, respectively.}
\label{H_av}
\includegraphics[scale=0.6]{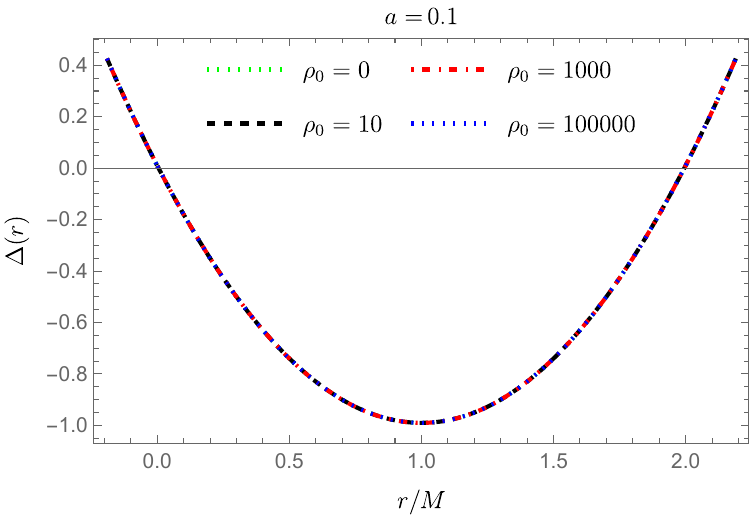}
\includegraphics[scale=0.6]{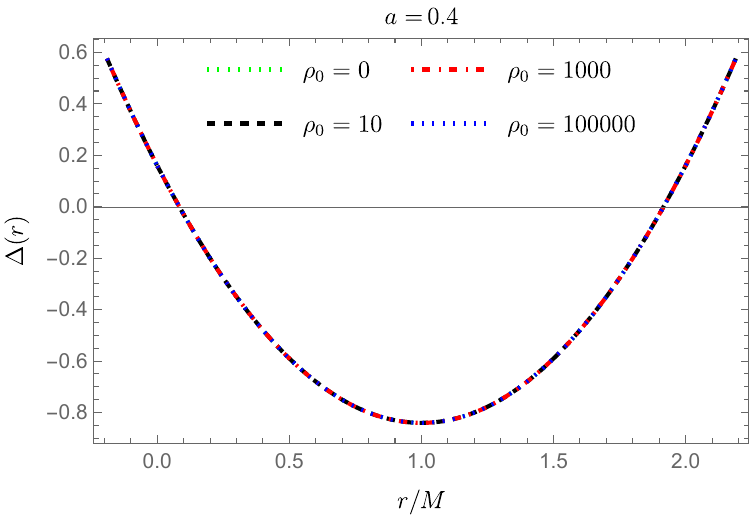}
\includegraphics[scale=0.6]{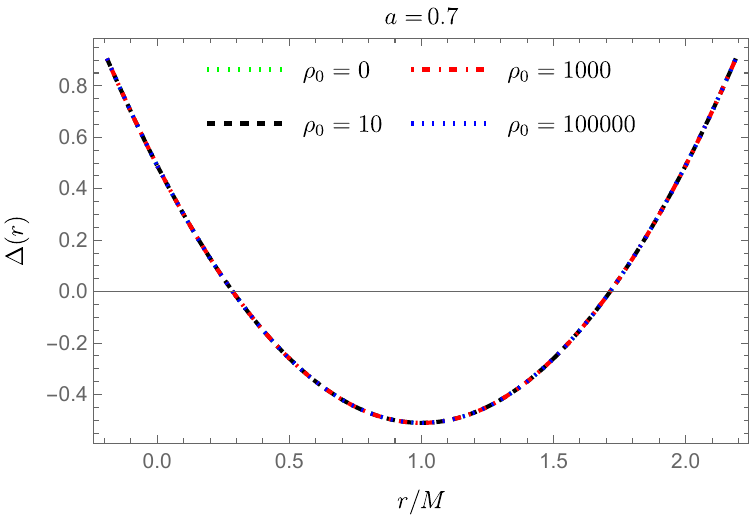}
\includegraphics[scale=0.6]{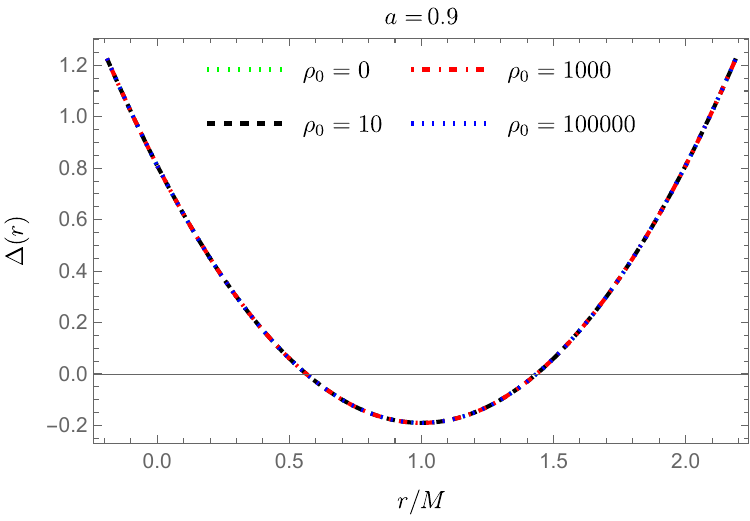}
\setlength{\abovecaptionskip}{-0.3cm}
\setlength{\belowcaptionskip}{0.9cm}
\caption{$\Delta(r)$ as the function of $r$ (see Eq. (\ref{DeltaGL})), with $r_0=91.2 \mathrm{kpc}$. From left to right and from top to bottom the values of $a$ are $a=0.1$, $a=0.4$, $a=0.7$ and $a =0.9$ result, respectively. Here we study the effect of different dark matter densities on the event horizon. The green curve represents $\rho_0=0M_{\odot} / k p c^3$, the black curve represents $\rho_0=10M_{\odot} / k p c^3 $, the red curve represents $\rho_0=1000M_{\odot} / k p c^3$, and the blue curve represents $\rho_0=100000M_{\odot} / k p c^3$.}
\label{H_rhov}
\end{figure}

The Kerr black hole in general relativity usually has two event horizons, the outer event horizon and the inner event horizon. For the rotating black hole surrounded by the pseudo-isothermal halo we obtained in the previous section, it is worth studying whether its event horizon is affected by dark matter. For a Kerr-like black hole, the event horizon is obtained by solving the equation $\Delta(r)=0$. For the rotating black hole surrounded by the pseudo-isothermal halo, the event horizon is obtained by solving the following equation
\be
\Delta(r)=\left(r_0^2+r^2\right){}^{4 \pi  \rho _0 r_0^2} \text{exp}\left[\frac{8 \pi  \rho _0 r_0^3  \arctan
   \left(\frac{r}{r_0}\right)}{r}\right]-2 M r+a^2=0.
\ee
For this equation, we cannot give an analytical solution, so we use numerical analysis to study the event horizon structure. In Fig. \ref{H_av} and Fig. \ref{H_rhov}, we give $\Delta(r)$ as the function of $r$. From Fig. \ref{H_av} and Fig. \ref{H_rhov}, we can see that the rotating black hole surrounded by the pseudo-isothermal halo still has a inner event horizon and event horizon. From Fig. \ref{H_av}, one can find that as the spin parameter $a$ increases, the interval between the inner event horizon and the outer event horizon becomes smaller and smaller, and the rotating dark matter black hole finally becomes an extreme black hole. In Fig. \ref{H_rhov}, the green curve represents $\rho_0=0M_{\odot} / k p c^3$, the black curve represents $\rho_0=10M_{\odot} / k p c^3 $, the red curve represents $\rho_0=1000M_{\odot} / k p c^3$, and the blue curve represents $\rho_0=100000M_{\odot} / k p c^3$. From Fig. \ref{H_rhov}, one can see that the existence of dark matter has a tiny effect on the event horizon.

\section{Geodesics around the rotating black hole surrounded by the pseudo-isothermal dark matter halo}\label{HJ-eq}
It is well known that there are two main situations for the motion of photons propagating in a gravitational background of a black hole: i) photons with smaller impact parameters will fall into the black hole and cannot be detected; ii) photons with impact parameter greater than a certain critical value, and they can be detected by the observer. In this section, we will study the geodesic equation of the rotating spacetime surrounded by the pseudo-isothermal dark matter halo. Our main concern is the Hamilton-Jacobi equation, which is defined as
\begin{equation}
\frac{\partial \mathcal{I}}{\partial \lambda}=-\mathcal{H},
\end{equation}
where $\mathcal{I}$ denotes the Jacobi action, and $\lambda$ denotes the geodesic affine parameter. Furthermore, the Hamiltonian $\mathcal{H}$ reads
\begin{equation}
\mathcal{H}=\frac{1}{2} g^{\mu \nu} \frac{\partial \mathcal{I}}{\partial x^\mu} \frac{\partial \mathcal{I}}{\partial x^\nu}\,.
\end{equation}
For the rotational black hole  surrounded by the pseudo-isothermal halo, the Hamilton-Jacobi equation is given by
\begin{equation}\label{HJ}
\frac{\partial \mathcal{I}}{\partial \lambda}=-\frac{1}{2} g^{\mu \nu} \frac{\partial \mathcal{I}}{\partial x^\mu} \frac{\partial \mathcal{I}}{\partial x^\nu}.
\end{equation}
By using the method of the separation of variables, the action $\mathcal{I}$ assumes the form
\begin{equation}
\mathcal{I}=\frac{1}{2} \mu^2 \lambda-E t+L \phi+S_r(r)+S_\theta(\theta)
\end{equation}
where $\mu$ is the mass of the particle. $E$ and $L$ are the conserved quantities in the motion of photons along the geodesic. Moreover, $S_r(r)$ and $S_\theta(\theta)$ denote the radial and angular functions, respectively. Substituting the action $\mathcal{I}$ into Eq. (\ref{HJ}), and using the functions $\{S_r(r), S_\theta(\theta)\}$ and the conserved quantities $\{E, L\}$, we obtain
\begin{equation}\label{4.5}
\begin{aligned}
\Sigma \frac{d t}{d \lambda} & =E\left[\frac{\left(a^2+r^2\right)\left(a^2+r^2-a L/E\right)}{\Delta(r)}-a\left(a \sin ^2 \theta-L/E\right)\right] , \\
\Sigma \frac{d r}{d \lambda} & =\sqrt{R(r)}, \\
\Sigma \frac{d \theta}{d \lambda} & =\sqrt{\Theta(\theta)}, \\
\Sigma \frac{d \phi}{d \lambda} & =E\left[\frac{a\left(a^2+r^2-a L/E \right)}{\Delta(r)}-\left(a-L/E \csc ^2 \theta\right)\right],
\end{aligned}
\end{equation}
where $R(r)$ and $\Theta(\theta)$ are defined as
\begin{equation}\label{4.6}
\begin{aligned}
R(r)&=\left[E(r^2+a^2)-a L\right]^2-\Delta(r)\left[\mathcal{Q}+(aE-L)^2+\mu^2r^2\right],\\
\Theta(\theta)&=\mathcal{Q}-a^2( \mu^2-E^2)\cos ^2 \theta-L^2 \cot ^2 \theta,
\end{aligned}
\end{equation}
and $\mathcal{Q}$ represents the Carter constant \cite{Carter:1968rr}, which is an integral constant introduced by Carter.
The equations (\ref{4.5}) and (\ref{4.6}) are complete time-like geodesics and null geodesics equations. When $\mu=1$ they describe the motion of time-like particles, and when $\mu=0$ they describe the motion of null particles. For the black hole shadow, we consider null geodesics, i.e. $\mu=0$.
Note that $R(r)$ and $\Theta(\theta)$ must be greater than 0, i.e.
\begin{equation}
\frac{R(r)}{E^2}=[r^2+a^2-a \xi]^2-\Delta(r)\left[\eta+(\xi-a)^2\right] \geq 0,
\end{equation}
\begin{equation}
\frac{\Theta(\theta)}{E^2}=\eta+(\xi-a)^2-\left(\frac{\xi}{\sin \theta}-a \sin \theta\right)^2 \geq 0.
\end{equation}
where the impact parameters are defined as
\be
\xi=L/E, \quad
\eta=\mathcal{Q}/E^2.
\ee
The condition satisfied by the stable motion orbit region of the photon is  $R^{\prime\prime}(r)>0$, and the condition satisfied by the unstable motion orbit region of the photon is  $R^{\prime \prime}(r)<0$.
The critical orbit between the stable orbit and the unstable orbit satisfies the condition
\begin{equation}
\left.R(r)\right|_{r_{ph}}=0, \quad \left.\frac{\mathrm{d} R(r)}{\mathrm{d} r}\right|_{r_{ph}}=0,
\end{equation}
where $r_{ph}$ denotes the radius of the photo-sphere. According to the above  conditions, the critical impact parameters can be written as
\begin{equation}
\xi=\frac{a^2+r^2}{a}-\frac{4 r \left[a^2+r^2 \mathcal{A}_{PI}(r)-2 M r\right]}{a \left[r^2 \mathcal{A}_{PI}'(r)+2 r \mathcal{A}_{PI}(r)-2 M\right]},
\end{equation}
and
\begin{equation}
\begin{aligned}
\eta=\frac{r^5 \left[8 a^2 \mathcal{A}_{PI}'(r)+\left(2 \mathcal{A}_{PI}(r)-r \mathcal{A}_{PI}'(r)\right) \left(r^2 \mathcal{A}_{PI}'(r)-2 r \mathcal{A}_{PI}(r)+12 M\right)\right]+4 M
   r^3 \left(4 a^2-9 M r\right)}{a^2 \left[r^2 \mathcal{A}_{PI}'(r)+2 r \mathcal{A}_{PI}(r)-2 M\right]^2},
\end{aligned}
\end{equation}
where $\mathcal{A}_{PI}(r)$ is given by the equation  (\ref{fPI}), and $\mathcal{A}^{\prime}_{PI}(r)$ denotes the derivation $\frac{d\mathcal{A}_{PI}(r)}{dr}$.

\section{The shadow shape of rotating BH surrounded by the pseudo-isothermal halo}\label{shape_shadow}
Through the research in the previous section we obtained the geodesics of photons, and based on these results we can calculate the motion of photons detected in a rotating black hole surrounded by the pseudo-isothermal halo for any observer. To draw shadow images, people usually use celestial coordinates ($x$ and $y$), which is a two-dimensional coordinate system. For an observer at the position ($r,\theta$), the celestial coordinates are defined as
\cite{Johannsen:2013vgc,Hioki:2009na}
\begin{figure}[htbp]
\begin{center}
\vspace{0.5cm}
\includegraphics[scale=0.6]{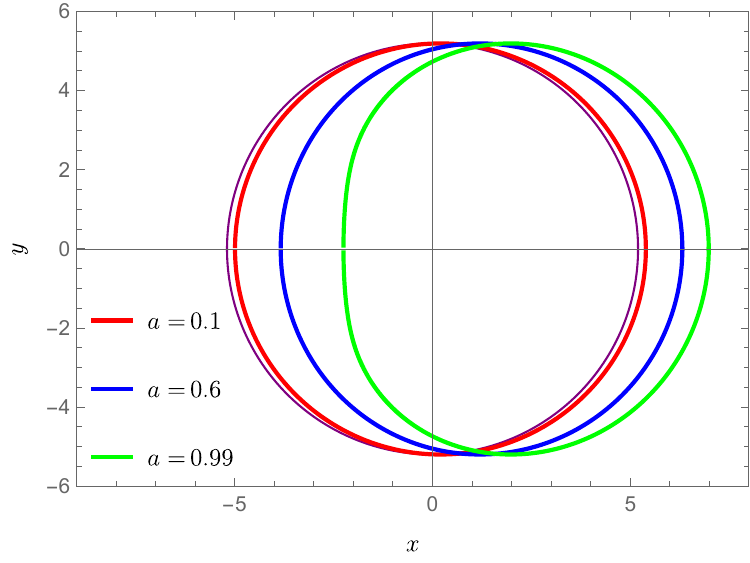}
\includegraphics[scale=0.6]{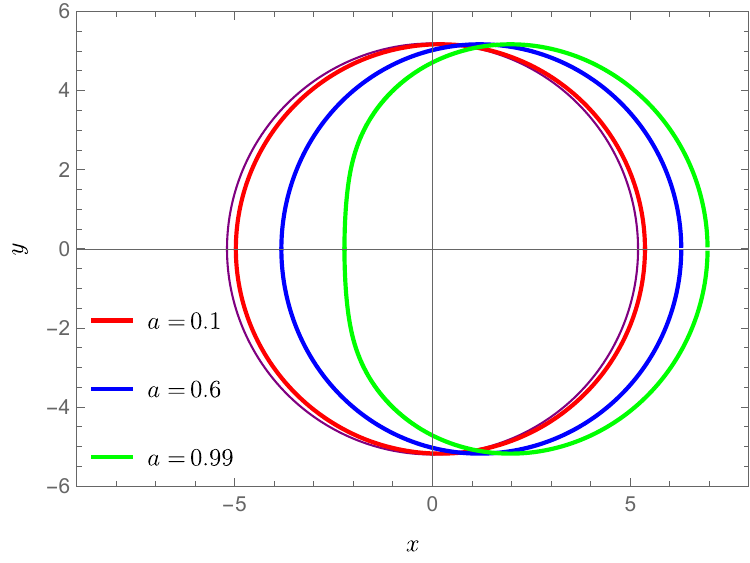}
\end{center}
\setlength{\abovecaptionskip}{-0.5cm}
\setlength{\belowcaptionskip}{0.3cm}
\caption{The shadow of the black hole as the spin parameter $a$ varies. The left panel is the shadow of the Kerr black hole when the dark matter density $\rho_0$ is zero. The right panel is the shadow of a rotating black hole surrounded by the pseudo-isothermal dark matter halo, where the parameters are set to $M=1, \rho_0=6.9 \times 10^6 M_{\odot} / k p c^3, r_0=91.2 \mathrm{kpc}$. Moreover, the purple solid line is the shadow of the Schwarzschild black hole.}
\label{av}
\includegraphics[scale=0.6]{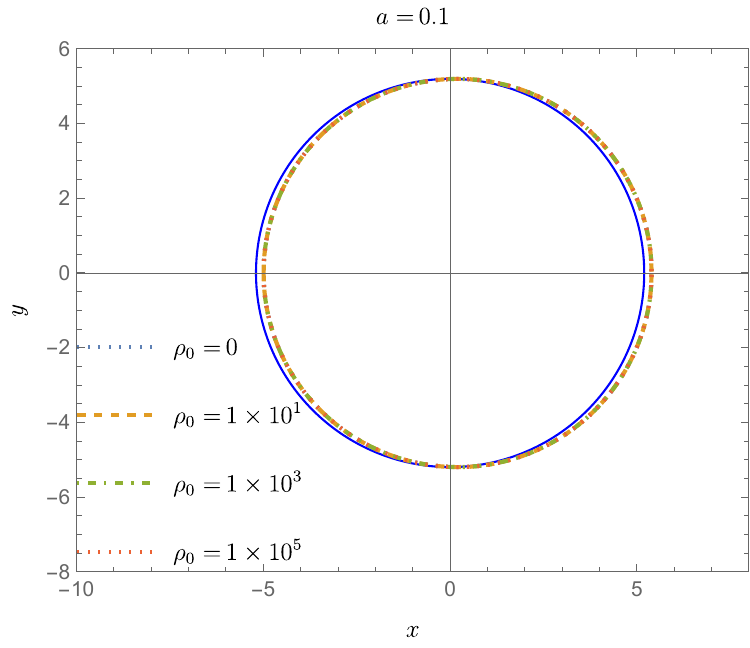}
\includegraphics[scale=0.6]{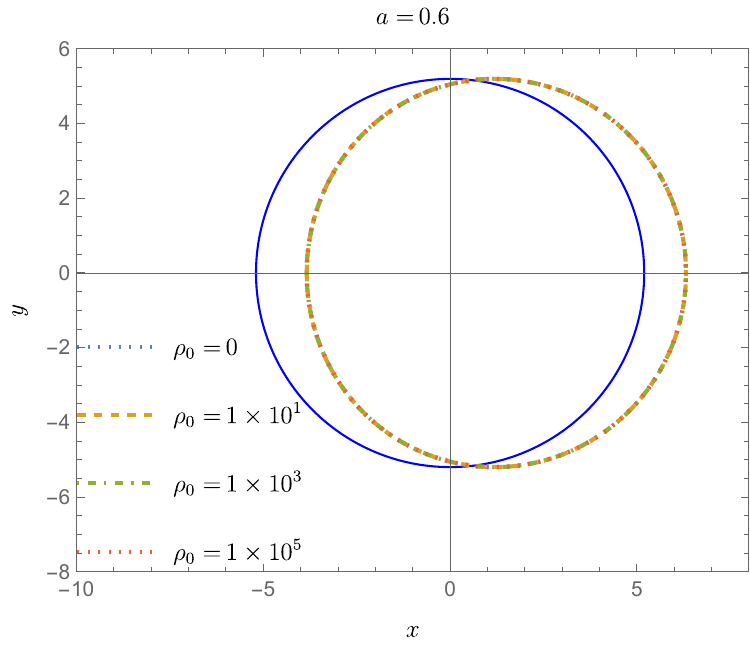}
\includegraphics[scale=0.6]{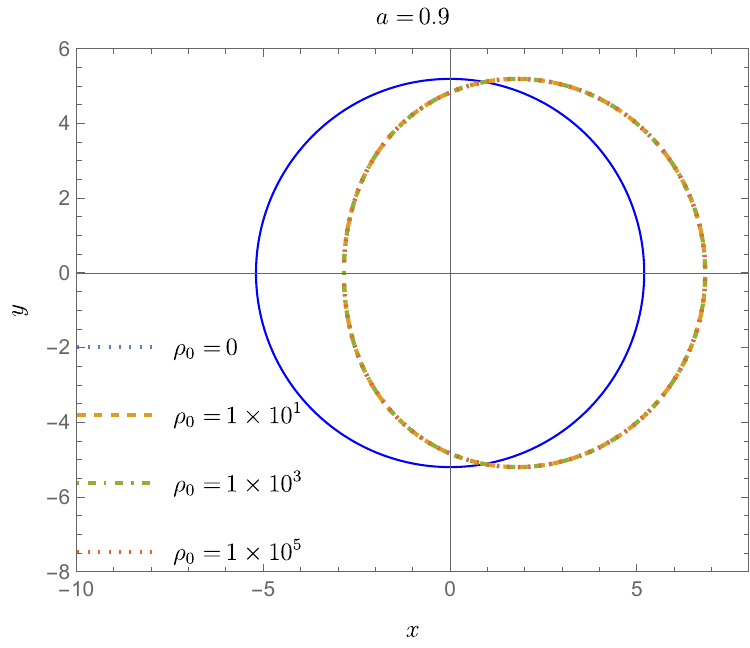}
\includegraphics[scale=0.6]{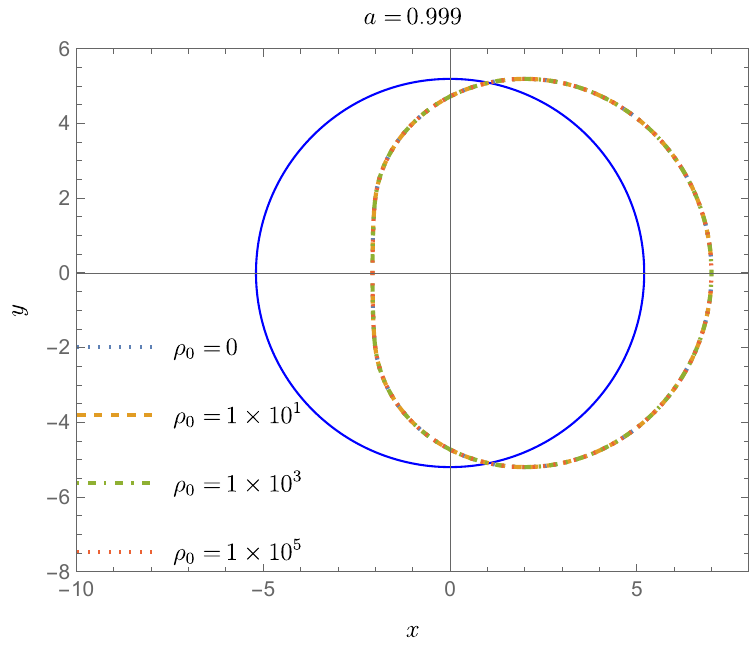}
\setlength{\abovecaptionskip}{-0.5cm}
\setlength{\belowcaptionskip}{0.3cm}
\caption{The shadow of the black hole as the different dark matter densities $\rho_0$ varies. From left to right and from top to bottom are different spin parameters $a=0.1$, $a=0.6$, $a=0.9$, and $a=0.999$. The blue solid line is the shadow of the Schwarzschild black hole, and $\rho_0=0$ corresponds to the shadow of the Kerr black hole. The other parameters are set to $M=1, r_0=91.2 \mathrm{kpc}$.}
\label{rhov}
\end{figure}
\be
\begin{aligned}
x &=\lim _{r \rightarrow \infty}\left(-r^2 \sin \theta \frac{d \phi}{d r}\right),\\
y &=\lim _{r \rightarrow \infty}\left(r^2 \frac{d \theta}{d r}\right).
\end{aligned}
\ee
where $r$ represents the space distance from the observer to the rotating black hole surrounded by the pseudo-isothermal halo, while $\theta$ represents the angle between the straight line determined by the observer and the center of the black hole, and the rotation axis of the black hole. The motion of photons is described by $d\phi/dt$ and $d\theta/dt$.
For asymptotically flat black hole metrics, the celestial coordinates can be simplified as
\be
x=-\frac{\xi}{\sin \theta},
\ee
and
\be
y= \pm \sqrt{\eta+a^2 \cos ^2 \theta-\xi^2 \cot ^2 \theta}.
\ee
In our work, we consider that the observer is located on the equatorial plane of the rotating black hole surrounded by the pseudo-isothermal halo, i.e., $\theta=\pi/2$. Therefore, the celestial coordinates can be written as
$
\label{alpha32}
x=-\xi$, and $y=\pm \sqrt{\eta}$.

Using $x$ and $y$ we can present the shadow image of the rotating black hole surrounded by the pseudo-isothermal halo. It allows us to study the influence of the presence of dark matter on the black hole through the image of the shadow. In Fig. \ref{av} we investigate the effect of the spin parameter $a$ on the shadow of the rotating black hole surrounded by the pseudo-isothermal halo. The left panel of Fig. \ref{av} is the shadow of the Kerr black hole when the dark matter density $\rho_0$ is zero. The right panel of Fig. \ref{av} is the shadow of rotating black hole surrounded by the pseudo-isothermal dark matter halo, where the parameters are set to $ M=1, \rho_0=6.9 \times 10^6 M_{\odot} / k p c^3, r_0=91.2 \mathrm{kpc}$. Moreover, the purple solid line is the shadow of the Schwarzschild black hole, which is a standard circle. From Fig. \ref{av}, one can find that as the spin parameter $a$ increases, the shadow of the rotating black hole surrounded by the pseudo-isothermal halo gradually deviates from a standard circle. The left half of the shadow image is gradually distorted, while the right half of the shadow is still a standard semicircle. Furthermore, the shadow of the rotating black hole surrounded by the pseudo-isothermal halo is very similar to that of the Kerr black hole.

In Fig. \ref{rhov} we study the effect of dark matter density on the black hole shadow. From left to right and from top to bottom in Fig. \ref{rhov} are different spin parameters $a=0.1$, $a=0.6$, $a=0.9$, and $a=0.999$. The blue solid line is the shadow of the Schwarzschild black hole, and $\rho_0=0$ corresponds to the shadow of the kerr black hole. The other parameters are set to $M=1, r_0=91.2 \mathrm{kpc}$. From Fig. \ref{rhov}, one can find that the dark matter density has a very little influence on the black hole shadow.

\section{Conclusion} \label{sec:summary}

The abundance of dark matter in our universe is substantial, leading us to consider a black hole enveloped by dark matter. Our research involves a link of the spherically symmetric Schwarzschild black hole with the pseudo-isothermal halo profile, resulting in a spherically symmetric black hole metric surrounded by dark matter. Observations of black hole photos taken by EHT and gravitational wave events have shown that black holes likely possess spin. To account for this, we employ the Newman-Janis method to derive a rotating black hole surrounded by the pseudo-isothermal halo, which brings us closer to a black hole solution akin to those in the real universe.

Our investigation delves into the effects of the pseudo-isothermal halo profile on the black hole event horizon, time-like and null orbits, and the shadow of the rotating black hole. We find that as the spin parameter $a$ increases, the interval between the inner event horizon and the outer event horizon of the rotating black hole surrounded by the pseudo-isothermal halo in M87 diminishes, yielding eventually the formation of an extreme black hole. Moreover, our research reveals that the presence of dark matter has minimal impact on the event horizon. However, as the spin parameter $a$ increases, the shadow of the rotating black hole surrounded by the pseudo-isothermal halo in M87 deviates increasingly from a standard circle, with larger spin parameters causing more pronounced distortion relative to the standard circle.

Interestingly, the dark matter density appears to have very little influence on the black hole shadow.
The shadows we observe suggest that dark matter density does not significantly affect the black hole shadow, even with varying densities. The observed shadow images closely resemble those of the Kerr black hole, even in the presence of dark matter, indicating that dark matter has minimal influence on the shadow. This raises the question of whether gravitational waves emitted by the rotating black hole surrounded by the pseudo-isothermal halo would also be challenging to distinguish from those of the Kerr black hole. A relatively important stage in the gravitational wave signal is the quasinormal mode, and it carries the unique characteristics of the background spacetime. Therefore, it is necessary to further study the quasinormal mode of these black holes surrounded by the pseudo-isothermal halo.
We intend to explore the quasinormal mode of the rotating black hole surrounded by the pseudo-isothermal halo in M87, which will be presented in a future work. The insights gained from gravitational waves and shadow observations hold the potential to guide experimental detection efforts.

\begin{acknowledgments}
This research was funded by the National Natural Science Foundation of China (No. 12265007 and 12261018), Universities Key Laboratory of System Modeling and Data Mining in Guizhou Province (No.2023013), the Science and Technology Foundation of Guizhou Province (No. ZK[2022]YB029). The work of G.L.  is supported by the Italian Istituto Nazionale di Fisica Nucleare (INFN) through the ``QGSKY'' project and by Ministero dell'Istruzione, Universit\`a e Ricerca (MIUR). G.L.,  and A. {\"O}.  would like to acknowledge networking support by the COST Action CA18108 - Quantum gravity phenomenology in the multi-messenger approach (QG-MM). A. {\"O}. would like to acknowledge the contribution of the COST Action CA21106 - COSMIC WISPers in the Dark Universe: Theory, astrophysics and experiments (CosmicWISPers) and the COST Action CA22113 - Fundamental challenges in theoretical physics (THEORY-CHALLENGES).

\end{acknowledgments}

\bibliography{DBH_ref}
\bibliographystyle{reference}
\end{document}